# Universal recovery of the bright-exciton level-degeneracy in quantum dots without structural symmetry

R. Trotta[1*], E. Zallo[1], C. Ortix[2], P. Atkinson[1,3], J. D. Plumhof[1], J. van den Brink[2], A. Rastelli[1,4], and O. G. Schmidt[1]

[1] *Institute for Integrative Nanosciences, IFW Dresden, Helmholtzstr. 20, D-01069 Dresden, Germany*

[2] *Institute for Theoretical Solid State Physics, IFW Dresden, Helmholtzstr. 20, D-01069 Dresden, Germany*

[3] *Institut des NanoSciences des Paris, UPMC CNRS UMR 7588, 4 Place Jussieu Boite courier 840, Paris 75252 Cedex 05, France*

[4] *Institute of Semiconductor and Solid State Physics, Johannes Kepler University Linz, Altenbergerstr. 69, A-4040 Linz, Austria*

The lack of structural symmetry which usually characterizes semiconductor quantum dots lifts the energetic degeneracy of the bright excitonic states and hampers severely their use as high fidelity sources of entangled photons. We demonstrate experimentally and theoretically that it is always possible to restore the excitonic degeneracy by the simultaneous application of large strain and electric fields, despite the fact that this possibility has fundamentally been doubted. This is achieved by using one external perturbation to align the polarization of the exciton emission along the axis of the second perturbation, which then erases completely the energy splitting of the states. This result, which holds for any QD structure, highlights the potential of combining complementary external fields to create artificial atoms meeting the stringent requirements posed by scalable semiconductor-based quantum-technology.

*electronic mail: r.trotta@ifw-dresden.de

Semiconductor quantum dots (QDs) are nanostructures made of several thousands of atoms which can self-assemble during epitaxial growth. In spite of the remarkable progress in QD fabrication [1, 2] and even in the ideal case in which defects and impurities can be neglected, unavoidable fluctuations in the exact number of atoms, their arrangement in the host matrix and intermixing with the substrate and the cap material make each QD unique. As a result, the possibility to grow semiconductor QDs showing specific properties of symmetry is a mere theoretical construct [3, 4]. This represents an enormous obstacle towards the exploitation of QDs in quantum technologies, where the precise control over the QD emission properties is a fundamental requirement [5]. In particular, the absence of structural symmetry usually occurring in real QDs induces a coherent coupling of the two bright excitonic states [6-10] which leads to an energetic separation between them - the so-called fine structure splitting (*FSS*). When the *FSS* is larger than the radiative linewidth of the transitions (of the order of 1 μeV), the fidelity of the entangled photon pairs emitted during the decay biexciton-exciton-ground state is strongly reduced [11-14]. The origin and magnitude of the *FSS* are mainly determined by the QD morphology, piezoelectricity, alloying and the symmetry of the underlying lattice [15-18]. For more than a decade researchers have struggled to find a reproducible way to suppress the *FSS* and, to date, some quantum optics experiments are still carried out with the rare QDs showing small *FSS* [19, 20]. Recently, the possibility to use external perturbations such as electric [6, 21], magnetic [14], and strain fields [7, 22] to erase the *FSS* has been explored. Theoretical [23, 24] and experimental [6, 7] results have demonstrated that there is a lower bound of the *FSS* (usually larger than 1 μeV) caused by the coherent coupling of the two bright exciton states. On the other hand, there are a few works reporting values of the *FSS* close to zero [14, 25], but it remains unclear whether a strictly zero *FSS*=0 can be universally achieved, especially considering that it is commonly believed that the excitonic degeneracy cannot be restored in QDs with low structural symmetry [23, 26]. In this work, we demonstrate for the first time that the coupling between the bright excitons can be reversibly controlled and erased by the simultaneous application of large strain and electric fields, thereby showing that it is always possible to drive the excitons confined in arbitrary QDs towards a universal level crossing.

The device employed in this work is shown schematically in Fig. 1(a). Thin p-i-n nanomembranes containing QDs are integrated onto $[Pb(Mg_{1/3}Nb_{2/3})O_3]_{0.72}$-$[PbTiO_3]_{0.28}$ (PMN-PT) piezoelectric actuators by gold-thermocompression bonding, as described elsewhere [27]. The thickness of the nanomembrane and the position of the QD layer were optimized to enhance the light-extraction-efficiency via a simple metal-semiconductor planar cavity [27]. The quantum emitters (InGaAs QDs grown on GaAs (001)) were grown by solid source molecular beam epitaxy and were placed in the centre of a 10 nm-thick GaAs/$Al_{0.4}Ga_{0.6}$As quantum well. This layer sequence reduces carrier ionization at high electric fields across the diode ($F_d$) and allows the QD emission lines to be shifted over a broad spectral range by applying a reverse bias ($V_d$) to the nanomembranes [6]. Simultaneously, a voltage $V_p$ (electric field $F_p$) applied to the PMN-PT induces anisotropic in-plane (compressive or tensile) biaxial strains in the QD layer [28], which allow the energy of the QD emission lines to be controlled in a well-defined manner [27, 29].

Micro-photoluminescence spectroscopy was used for the optical characterization of the device. The measurements were performed at low temperature (typically 4-10 K) in a helium flow cryostat and with a microscope objective with numerical aperture of 0.42. The device was excited with a 850 nm continuous-wave Ti:Sa laser and the signal was collected by the same microscope objective used for the excitation, spectrally analyzed by a single or double spectrometer with 0.75 m focal length and detected by a Si charge-coupled device. Polarization-resolved measurements were performed by rotating a half lambda waveplate in combination with a linear polarizer, whose transmission axis was set along the [110] crystal direction of the nanomembranes with an accuracy of 10°. The *FSS* and the polarization angle of the excitonic emission were evaluated using the same procedure reported in Ref. 6, which ensures sub-microelectronvolt resolution.

The device shown in Fig. 1(a) combines the most powerful "tuning knobs" available to date (electric and strain fields) and allows the QD emission properties to be engineered on demand. This is for example demonstrated in Fig. 1(b), where the two knobs are used for tuning all the QD emission lines across a spectral range larger than 35 meV, which is one of the largest reported so far. Furthermore, control over the binding energies of biexciton (XX) and charged excitons ($X^+$ and $X^-$) is also demonstrated in Fig. 1(b), as reported previously with either electric fields or strain [6, 29]. We now show that the possibility to apply *simultaneously* two *independent* external fields allows a fundamental problem plaguing QDs – the presence of a coherent coupling of excitons – to be solved.

The coupling between the bright excitonic states can be highlighted by sweeping their energetic levels one across the other [6, 7, 23]. This is shown schematically in Fig. 2(a), where the two excitons ($X_1$ and $X_2$) undergo an avoided crossing as the magnitude of an external field is varied. The minimum *FSS* value ($\Delta$) is a measure of the coupling strength and $\theta$ is the polarization angle of one of the transitions with respect to the [110] crystal direction. Figure 2(b)-(c) show the experimental behaviour of *FSS* and $\theta$ for a QD as a function of $F_p$ ($F_d$) with $V_d$=0 ($V_p$=0). It is evident that the effects produced by the two fields on *FSS* and $\theta$ are *qualitatively* similar (they follow the anticrossing pattern sketched in Fig. 2(a)), but *quantitatively* distinct. In particular, the values of $\Delta$ are very different, implying that the coupling is not only an intrinsic property of the QD under consideration, but it is influenced by the strength and symmetry of the external perturbations.

Figure 3(a) shows the behaviour of the *FSS* for another QD as a function of $F_d$ for different values of $F_p$. For tensile strain ($F_p$<0), the value of $\Delta$ increases up to a maximum observed value of $\Delta_{max}$~25 µeV, while under compressive strain ($F_p$>0) $\Delta$ first decreases to a minimum value ($\Delta_{min}$) and then increases again. As can be observed more clearly in Fig. 3(b) where the *FSS* is plotted against the exciton energy ($E_x$), the value of $\Delta_{min}$ is comparable with the experimental spectral resolution (~0.5 μeV). Similar results were obtained in all QDs chosen randomly in our device (see Fig. 3(c)), where it is always possible to tune the FSS down to the experimental spectral resolution. This evidence, along with the fact that even the *FSS* of a QD showing $\Delta_{max}$ as large as 40 μeV can be tuned to values below 0.5 μeV, strongly suggests that electro-elastic fields enable us to "erase" the *FSS* of most of (if not all) the QDs in our sample. Additional information can be found studying the behaviour of $\theta$ (see Fig. 4(a)) for the same QD reported in Fig. 3(a). When $F_p$ is increased, the $\theta$ curve shows sharper variations with $F_d$ and, after the critical point where $\Delta=\Delta_{min}$, the direction of rotation is inverted, from anticlockwise to clockwise. This change in the handedness of $\theta$ has never been observed in the same QD [6], and suggests that the *FSS*=0 critical point has been crossed. Obviously, the finite spectral resolution of the experimental apparatus prevents *FSS*=0 to be proved but the fact that we observe systematically the same behaviour in all the QDs we measured indicates the existence of a universal behaviour of the *FSS* under the influence of strain and electric fields, despite the different properties of the quantum emitters. In order to check if *FSS*=0 can be achieved we now analyze the experimental observations theoretically.

The *FSS* originates from the breaking of the $D_{2d}$ or $C_{4v}$ symmetry occurring even in ideal QDs with $C_{2v}$ symmetry [10]. Any additional disorder – we can safely assume that any realistic QD has only $C_1$ symmetry – implies that the polarization angle of the bright excitons further departs from the pure lens shaped ones [10, 24, 30]. Whether or how the *FSS* can be tuned to zero depends on the existence of external perturbations able to "universally" restore the level degeneracy, independent of the specific QD parameters. By considering the combined effect of a vertical electric field (*F*) applied along the [001] direction and anisotropic biaxial stresses [28] of magnitude $p = p_1 - p_2$, where $p_1$ and $p_2$ are the magnitudes of two perpendicular stresses applied along arbitrary directions in the (001) plane, the effective two-level Hamiltonian for the bright excitons takes the form (see Supplementary Material):

$$H = [\eta + \alpha p + \beta F]\sigma_z + [k + \gamma p]\sigma_x,$$

where $\sigma_{z,x}$ are the usual Pauli matrices, $\eta$ and $k$ accounts for the QD structural asymmetry. The parameters related to the external fields are $\alpha$ and $\gamma$ (related to the elastic compliance constants renormalized by the valence band deformation potentials), and $\beta$ (proportional to the difference of the exciton dipole moments). Diagonalization of the above Hamiltonian gives the following values of *FSS* and $\theta$:

$$FSS = \left[(\eta + \alpha p + \beta F)^2 + (k + \gamma p)^2\right]^{1/2}, \tag{1}$$

$$\tan\theta_\pm = \frac{k + \gamma p}{\eta + \alpha p + \beta F \pm FSS}. \tag{2}$$

Equations (1)-(2) are used to fit the experimental data with the parameters of the QD fixed (see Supplementary Material). As shown in Fig. 3(a) and Fig. 4(a) an excellent agreement is found. Even more, the theory predicts *FSS*=0 and, after this critical point, there is an inversion in the handedness of $\theta$. Therefore, the change in handedness can be considered as the experimental signature of the crossing of the FSS=0 critical point. Most importantly, it can be easily shown that equation (1) has *always* a minimum at zero when the magnitude of *F* and *p* take the values

$$p_{critic} = -\frac{k}{\gamma}, \qquad F_{critic} = \frac{\alpha k}{\gamma\beta} - \frac{\eta}{\beta}.$$

In other words, there are *always* values of $F_p$ and $F_d$ such that *FSS*=0, regardless of the QD structure. It is obvious that in real experiments large enough tuning ranges are needed to access the values of $p_{critic}$ and $F_{critic}$ given above. This requirement is satisfied by our device, which allows us to tune systematically all the QDs we measure to *FSS*=0. It is also important to observe that the remarkable agreement between all the experimental data and the simple theory developed in this work indicates that higher order terms in *p* and *F* can be neglected. The result is in line with those of Ref. [24], which found good agreement between this continuum theory and atomistic calculations accounting for the microscopic effects that lower the QD symmetry to $C_1$ (note that Ref. 24 considered only the effect of a single perturbing field (strain), while here two independent fields are addressed).

Finally, by combining the theoretical analysis with the experimental data a clear and intuitive picture emerges. Figure 4(b) shows the dependence (in polar co-ordinates) of $\Delta E$ vs. $\varphi$ for specific values of $F_p$ and $F_d$, where $\Delta E = |E(\varphi,F_p,F_d)-E_{min}(F_p,F_d)|$ and $\varphi$ is the angle the polarizer forms with a rotating half lambda waveplate. $E(\varphi,F_p,F_d)$ is half of the difference between the exciton and biexciton emission (see Ref. 6) and $E_{min}(F_p,F_d)$ is the minimum value of $E(\varphi,F_p,F_d)$ averaged over $2\pi$ rotation of the waveplate. Therefore, the length and the orientation of the "petals" give the magnitude of the *FSS* and $\theta$, respectively. It is clear that when the eigenstates are oriented along the [110] (close to the [100]) or the perpendicular direction, the application of $F_d$ ($F_p$) leads to *FSS*=0. Since electric and strain fields act as effective deformations along, respectively, the [110] and close to the [100] directions (see Supplementary Material), this implies that a single perturbation (e.g. $F_d$) is sufficient to tune the *FSS* to zero only when one of the eigenstates is oriented along the direction of the perturbation, i.e., only when the external field is able to compensate *completely* the difference in the confining potentials of the two bright exciton eigenstates. For all other values of $\theta$, which are obtained for a non-optimum choice of the other field (e.g. $F_p \neq F_p^{crit}$), a lower bound of *FSS* is observed, in agreement with previous results obtained with a single tuning knob [6, 7, 23, 24]. The above discussion highlights the crucial importance of having at hand *two independent* and *broad-range* "tuning knobs" where we can use one perturbation (e.g. $F_p$) to align $\theta$ along a specific direction ([110]) and the other to tune *FSS* to 0.

In conclusion, we have presented a novel and all-electrically controlled device where the QD emission properties are engineered by large electro-elastic fields provided by diode-like nanomembranes integrated onto piezoelectric actuators. In particular, we have demonstrated that the simultaneous application of electric and elastic fields allow us for the first time to control and cancel the coherent coupling between the bright excitonic states – and hence the exciton fine structure splitting – in all the quantum dots we measure. The experimental observations are supported by a simple theoretical model which holds for every QD structure. The tedious search for QDs suitable for quantum optics experiments can therefore be avoided, and a deterministic implementation of QDs into a scalable semiconductor-based quantum-information technology seems much more likely in the foreseeable future.

**Acknowledgments**

We thank E. Magerl, S. Kumar, A. Schliwa, and V. Fomin for fruitful discussion, B. Eichler, R. Engelhardt, and D. Grimm for technical support, K. Dörr and A. Herklotz for help with the piezoelectric actuators. The work was supported financially by BMBF QuaHL-Rep (Contract n. 01BQ1032) and DFG FOR730.

**Figure Legends**

Fig. 1 (color online). (a), Sketch of the device. An n-i-p nanomembrane featuring the so-called "giant Stark effect" [6] is integrated on top of a piezoelectric actuator (PMN-PT) allowing the *in-situ* application of anisotropic biaxial strains by tuning the voltage (electric field) $V_p$ ($F_p$). Independently, a voltage applied to the nanomembrane ($V_d$) allows the electric field ($F_d$) across the QDs to be controlled, leading to a quantum confined Stark effect when the device is operating in reverse bias ($F_d<0$). (b), Colour-coded micro-photoluminescence (μ-PL) map of a single QD as a function of $F_p$ and $F_d$. The two fields were varied one after the other, i.e., $F_p$=-10 kV/cm when $F_d$ is ramped up (bottom), whereas $F_d$=-7 kV/cm when $F_p$ is ramped up (top). The abscissa indicates the energy of the emitted photons. The ordinate indicates the value of $F_p$ and $F_d$ used in the experiment. The exciton (X), biexciton (XX), and charged excitons ($X^+$ and $X^-$) transitions from the same QD were identified by polarization-resolved μ-PL spectroscopy.

Fig. 2 (color online). (a), Sketch of the energetic levels of the two bright excitonic states ($X_1$ and $X_2$) when they are swept across each other under the influence of a generic external field. The colours highlight the character of the states. The energetic distance between the two levels is the *FSS*, and $\Delta$ is a measure of the coupling strength between the states. The insets show a sketch of the angular distribution in the (001) plane of the light emitted by the two exciton transitions. The two components are assumed to have the same oscillator strength and to be oriented at 90° with respect to each other. The angle $\theta$ represents the polarization angle of the lowest energy transition with respect to the [110] crystal direction. (b), Measured behaviour (symbols connected by lines) of the *FSS* for a QD as a function of $F_p$ and $F_d$, top and bottom axis, respectively. (c), Behaviour of the polarization angle $\theta$ (symbols connected by lines) for the same QD of Fig. 2(b) as a function of $F_p$ and $F_d$, top and bottom axis, respectively. In (b) and (c), the data as a function of $F_p$ ($F_d$) are obtained with $V_d=0$ ($V_p=0$).

Fig. 3 (color online). (a), Behaviour of the *FSS* as a function of $F_d$. The different curves correspond to different values of $F_p$. Empty (full) circles show the experimental data for $F_p$ smaller (larger) than 13 kV/cm, which approximately corresponds to the critical strain condition for which the *FSS* reaches 0 as $F_d$ is varied. The solid lines are fits to the experimental data obtained using equations (1)-(2). $\Delta_{max}$ indicates the maximum value of $\Delta$ observed by combining both tuning knobs for this specific QD. (b), *FSS* as a function of the energy of the excitonic transition ($E_x$) in the region of small *FSS* (circles connected by lines). The colours correspond to the $F_p$ values shown in Fig. 3(a). The data taken at $F_p=12$ kV/cm, not displayed in Fig. 3(a) for clarity purposes, are also reported (orange points connected by lines). The value of the minimum $\Delta$ is denoted $\Delta_{min}$, which coincides with the experimental spectral resolution. (c), Histogram of $\Delta_{max}$ and $\Delta_{min}$ for the different measured QDs.

Fig. 4 (color online). (a), Behaviour of the polarization angle ($\theta$) as a function of $F_d$ for the same QD whose *FSS* data are reported in Fig. 3(a). Different curves correspond to different values of $F_p$. The same symbols and colours of Fig. 3(a) for the different values of $F_p$ have been used. The solid lines are fits to the experimental data obtained using equations (1)-(2). The direction of rotation of the eigenstates with $F_d$ is inverted after crossing *FSS*=0, which occurs for $F_p$~13 kV/cm. (b), Dependence (in polar coordinates) of $\Delta E$ vs. $\varphi$, where $\Delta E = |E(\varphi,F_p,F_d)-E_{min}(F_p,F_d)|$, $\varphi$ is the angle the polarizer forms with a rotating half lambda waveplate. The length and the orientation of the "petals" give the value of the *FSS* and $\theta$, respectively, see text. Specific values of $F_p$ and $F_d$ around *FSS*~0 have been used to construct this plot. Note that around the point where the *FSS*=0 (central panel) the phase rotates in the whole tuning range (180°) under the effect of the two fields.  According to Fig. 2(a), 0° and 90° correspond to the [110] and [1-10] crystal directions, respectively.

(a)

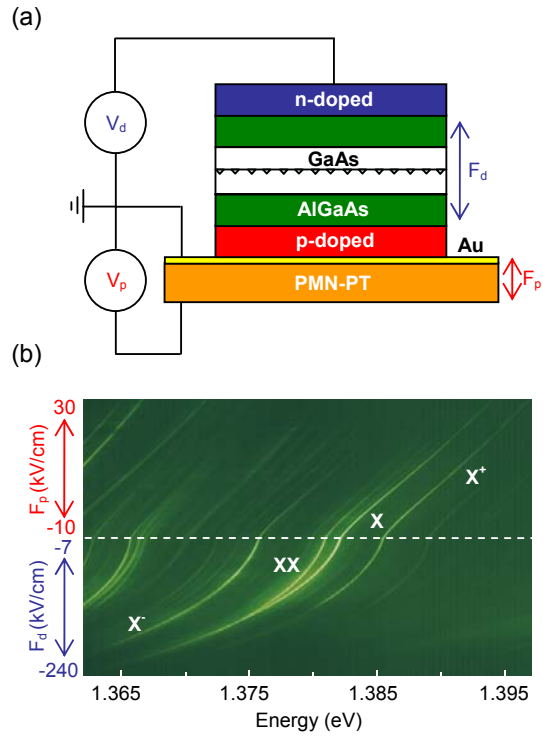

(b)

Figure 1 of 4

Universal recovery of the bright-exciton level-degeneracy in quantum dots without structural symmetry

by R. Trotta *et al.*

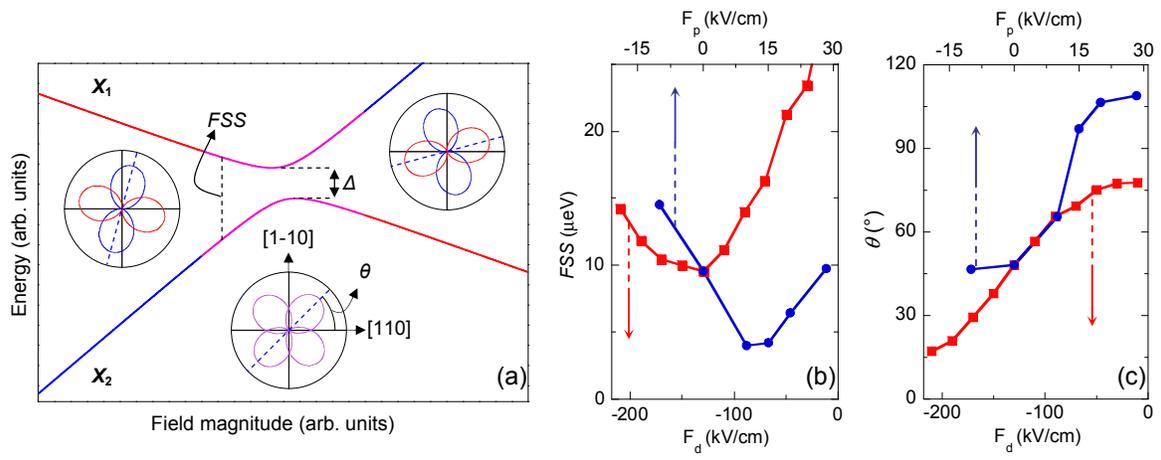

Figure 2 of 4

Universal recovery of the bright-exciton level-degeneracy in quantum dots without structural symmetry

by R. Trotta *et al.*

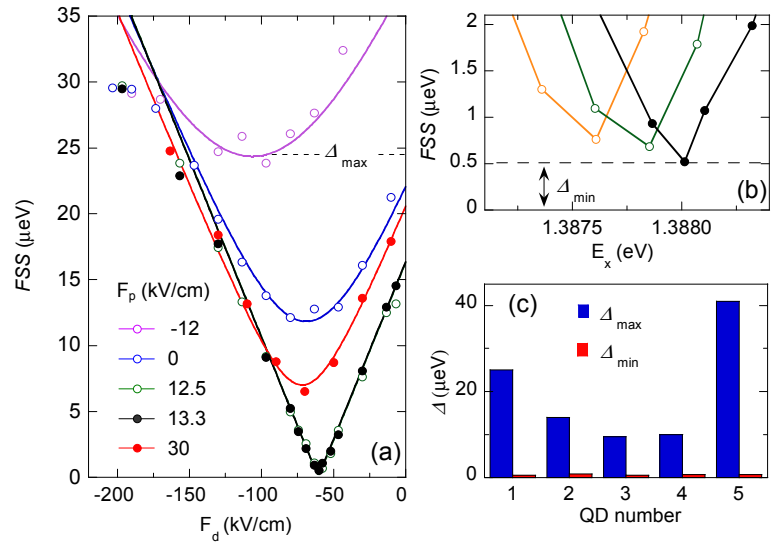

Figure 3 of 4

Universal recovery of the bright-exciton level-degeneracy in quantum dots without structural symmetry

by R. Trotta *et al.*

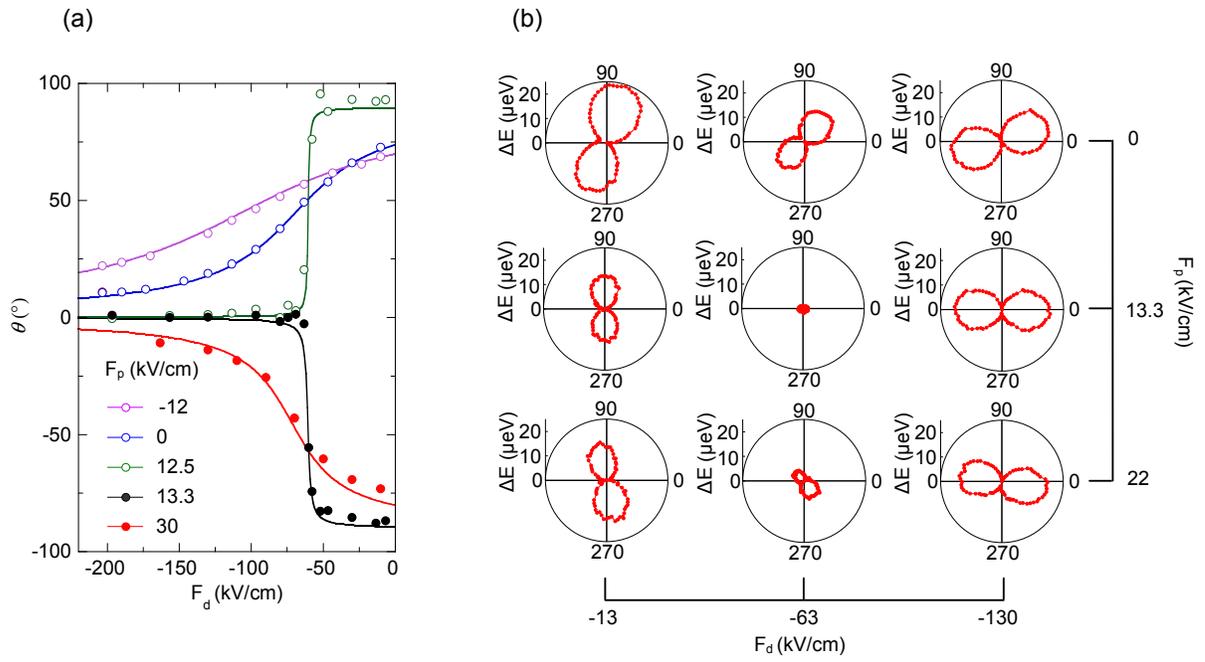

Figure 4 of 4

Universal recovery of the bright-exciton level-degeneracy in quantum dots without structural symmetry

by R. Trotta *et al.*